# Improved model for the soliton-potential interaction in $\varphi^4$ Model

**Ensieh Hakimi**
*en_ha92@stumail.um.ac.ir*
**Kurosh Javidan**

*javidan@um.ac.ir*

Department of physics, Ferdowsi university of Mashhad, 91775-1436 Mashhad Iran

**Abstract**
An improved model for the soliton-potential scattering is presented. This model is constructed with a better approximation for adding the potential to the lagrangian through the metric of background space-time. The results of the model are compared with other models and the differences are discussed.

## 1. INTRODUCTION

Scattering of solitons from potentials have been studied in many papers by different methods. It is shown that the soliton acts as a point-like particle in most of the cases. But there are some surprising features which need more attentions. Solitons interact on a potential barrier almost elastically; while they show quantum like behaviour during the interaction with a potential well. Several models have been presented for explaining such behaviours. Models are different in the method of adding the potential to the soliton equation of motion. The general behaviour of the solitons in all the models is the same, but details of the scattering and some interesting features are different. For example some of the models predict that a soliton may reflect backward (or forward) after the scattering on a potential well while other models have not such this behaviour. The potential can be added to the equation of motion as a perturbative term [1, 2]. These effects also can be taken into account by making some parameters of the equation of motion to be function of space or time [3]. Also one can add such effects to the Lagrangian of the system by introducing a suitable nontrivial metric for the back ground space-time, without missing the topological boundary conditions. This method has been used for studying the sine-Gordon model in [4, 5].

Scattering of $\varphi^4$ solitons on Barriers and Holes has been investigated with two different models in [6]. In one of the models, the potential has been considered by deforming one of the parameters of the soliton equation of motion. In another model the same potential has been added through the metric of background space-time. Some differences between the results of two models have been reported in this paper. Also there are some other important differences between the models which need attentions. There are some important questions: What is the reason for the differences between the results of two models? Which model makes a better explanation for the nature of the system? Is it possible to find a better approximation for one of the models (in fact the metric model) in order to have agreement between the models? If the answer is yes then we can conclude that two models are well. Note that these two models rise from different approaches. This investigation will give us a brighter situation from the soliton-potential scattering.

The answer is important from the other point of view. The results can show us which behaviours are model independent and completely come from the physics of solitons. Also we can find sharper knowledge about the effects of the modelling methods on the results of an investigation on nonlinear systems. The results also are very important for constructing suitable collective coordinate variables in the modelling of nonlinear systems.

The metric model has been used in constructing a collective coordinate system for topological solitons in [7]. Therefore if we find a better approximation for this model then we can improve the related models and their results.

Motivated by these questions we studied the similarities and the differences between the models. A brief description of the models and their results is presented in section 2. A better approximation for one of the models is presented in section3. New model is used for investigating the soliton-potential interaction in $\varphi^4$ model and the results are compared with the results of the other models in section 4. Some conclusions and remarks will be presented in section 5.

## 2. Two models for "$\varphi^4$ Solitons-potential" system

**Model 1**: Consider a scalar field with the lagrangian

$$\mathcal{L} = \frac{1}{2}\partial_\mu\varphi\partial^\mu\varphi - U(\varphi) \qquad (1)$$

and the following potential

$$U(\varphi) = \lambda(x)(\varphi^2 - 1)^2 \qquad (2)$$

The equation of motion for the field becomes

$$\partial_\mu\partial^\mu\varphi + 4\lambda(x)\varphi(\varphi^2 - 1) = 0 \qquad (3)$$

The effects of the potential are added to the equation of motion by using a suitable definition for $\lambda(x)$, like$\lambda(x) = 1 + v(x)$. For a constant value of parameter ($\lambda(x) = 1$) equation (2) has a solitary solution as following [6]

$$\varphi(x,t) = \pm\tanh\left(\frac{\sqrt{2}(x - x_0 - ut)}{\sqrt{1-u^2}}\right) \qquad (4)$$

in which $x_0$ and u are solitary wave initial position and its initial velocity respectively. This equation is used as initial condition for solving (3) with a space dependent $\lambda(x)$ when the potential v(x) is small.

**Model 2**: The potential also can be added to the lagrangian of the system, through the metric of background space-time. So the metric includes characteristics of the medium. The general form of the action in an arbitrary metric is:

$$I = \int \mathcal{L}(\varphi, \partial^\mu\varphi)\sqrt{-g}\,d^n x\, dt \qquad (5)$$

where "g" is the determinant of the metric $g^{\mu\nu}(x)$. A suitable metric in the presence of a weak potential v(x) is [4, 5 and 6]:

$$g^{\mu\nu}(x) \cong \begin{pmatrix} 1 + v(x) & 0 \\ 0 & -1 \end{pmatrix} \qquad (6)$$

The equation of motion for the field $\varphi$ which is described by the lagrangian (1) in the action (5) is [4, 7]:

$$\frac{1}{\sqrt{-g}}\left(\sqrt{-g}\partial_\mu\partial^\mu\varphi + \partial_\mu\varphi\partial^\mu\left(\sqrt{-g}\right)\right) + \frac{\partial U(\varphi)}{\partial\varphi} = 0 \qquad (7)$$

This equation of motion in the background space-time (6) becomes [6]:

$$\left(1 + v(x)\right)\frac{\partial^2\varphi}{\partial t^2} - \frac{\partial^2\varphi}{\partial x^2} - \frac{1}{2|1+v(x)|}\frac{\partial v(x)}{\partial x}\frac{\partial\varphi}{\partial x} + \frac{\partial U(\varphi)}{\partial\varphi} = 0 \qquad (8)$$

The field energy density is:

$$\mathcal{H}_2 = g^{00}(x)\left(\frac{1}{2}g^{00}(x)\dot{\varphi}^2 + \frac{1}{2}\acute{\varphi}^2 + U(\varphi)\right) \qquad (9)$$

The energy density is calculated by varying both the field and the metric (See page 643 equation (11.81) of [8]). If we look at the $\mathcal{L}(\varphi, \partial^\mu\varphi)\sqrt{-g}$ of action (5) as an effective lagrangial in a flat space-time, then the energy density becomes [6]

$$\epsilon = \sqrt{g^{00}(x)}\left(\frac{1}{2}g^{00}(x)\dot{\varphi}^2 + \frac{1}{2}\acute{\varphi}^2 + U(\varphi)\right) \text{(10)}$$

The above Hamiltonian density can be found by varying only the field in the effective lagrangian. This equation does not contain energy exchange between the field and the metric (therefore with the potential). This is an important point that makes some differences.

Solution (4) can be used as initial condition for solving (8) when the potential v(x) is small. If we define the parameter $\lambda(x) = 1 + v(x)$ in model 1 then it is possible to compare the results of two models. The potential

$$v(x) = \begin{cases} \lambda_0 & |x| < p \\ 0 & otherwise \end{cases} \qquad (10)$$

has been chosen in [4] where $\lambda_0$ and 'p' are the potential strength and the potential width respectively.

Scattering of a soliton with a potential barrier is nearly elastic. The soliton radiates a small amount of energy during the interaction. The radiated energy during the interaction in model 2 is more than the energy radiation in the model 1. The radiated energy becomes larger when the height of the barrier or the speed of the soliton increases [4].

Both two models show that there exist two different kinds of trajectories during the scattering of a soliton on a potential barrier depending on soliton initial velocity. Two kinds of trajectories are separated by a critical velocity $u_c$. At low velocities ( $u < u_c$) soliton reflects back and reaches its initial place. A soliton with an initial velocity $u > u_c$ has enough energy to pass over the potential.

Note that model 2 is valid for slowly varying small potentials [4]. It means that we have to use smooth and small potentials in metric (6), but a square-like potential is not smooth. Some simulations have been set up with a smooth potential $v(x) = a\,Exp(-b(x-c)^2)$ with using two models. Simulations show that the differences between the soliton behaviour in two models with using this potential decrease, but the differences do not vanish. Indeed the simulations with this potential are in agreement with the mentioned differences which we have been reported for a square-like potential in [4]. It means that the two models are really different in these cases.

Is it possible to improve a model in order to get agreement with the other model?

## 3. Improved approximation

It is possible to improve model 2 by using a better approximation for the metric of back ground space-time. The better metric is

$$g^{\mu\nu}(x) \cong \begin{pmatrix} 1+v(x) & 0 \\ 0 & -\frac{1}{1+v(x)} \end{pmatrix} \quad (11)$$

The field equation of motion is derived from (7) using the metric (11) as follows

$$\left(1+v(x)\right)\frac{\partial^2\varphi}{\partial t^2} - \frac{1}{1+v(x)}\frac{\partial^2\varphi}{\partial x^2} + \frac{\partial U(\varphi)}{\partial\varphi} = 0 \quad (12)$$

The energy density of the field in this improved model (which we call it 'model 3') becomes

$$\mathcal{H}_3 = \frac{1}{2}(1+v(x))^2\dot{\varphi}^2 + \frac{1}{2}\acute{\varphi}^2 + \left(1+v(x)\right)U(\varphi) \quad (13)$$

The potential is still a slowly varying and small function.

The models can be compared using the energy of a soliton in three models. The energy density of a soliton in model 1 is

$$\mathcal{H}_1 = \frac{1}{2}\dot{\varphi}^2 + \frac{1}{2}\acute{\varphi}^2 + \left(1+v(x)\right)U(\varphi) \quad (14)$$

The above energy density is related to the field only. Equations (9) and (13) are the energy density of " soliton + metric " in the model 2 and the new model3 respectively. Consider a static solitary wave located in the initial position $X_0$. The energy density of the field is $\frac{1}{2}\acute{\varphi}^2 + \left(1+v(x)\right)U(\varphi)$ in the models 1 and 3 while the model 2 predicts the static energy $\frac{1}{2}\left(1+v(x)\right)\acute{\varphi}^2 + \left(1+v(x)\right)U(\varphi)$ for this situation. The difference of the energy is $\Delta E_{static} = E_{model\ 3} - E_{model\ 1} = \frac{1}{2}v(x)\acute{\varphi}^2$. This means that the models 1 and 3 have the same effective potential but the model 2 contains an extra term. This small extra term is positive for a potential barrier and it is negative for potential well. It can be concluded that the effective potential in the model 2 is stronger than the effective potential of models 1 and 3. Let us now look at the kinetic terms in the three models. The kinetic energy in the models 2 and 3 are the same and they are different with the kinetic energy in the model 1. The difference kinetic energy in the first order approximation is $\Delta E_{kinetic} = E_{model\ 3} - E_{model\ 1} \cong v(x)\,\dot{\varphi}^2$. It is positive for potential barrier and negative for the potential well. This means that the effective mass in the model 1 is smaller than the effective mass of the soliton in the models 2 and 3 for a potential barrier. These differences act on the dynamics of solitary waves in the opposite ways therefore the predictions of the model 3 will be something between the predictions of the other models. The effects of these differences are studied in the next section.

## 4. Comparing the models

Several simulations using three models have been performed with different types of potentials. A potential of the form $v(x) = a\,Exp(-b(x-c)^2)$ has been used in the results which are presented below. This type of the potential is more suitable for the model 2 and 3 than a rectangular potential, because it is a smooth and slowly varying function.

Figure 1 shows the shape of the potential barrier $v(x) = 0.5\,Exp(-4x^2)$ as seen by the soliton in three models. The shape of this potential has been found by placing a static soliton at different positions and calculating its total energy. As we have seen, the static energy of the models 1 and 3 are equal while the static energy calculated using model 2 is different. Our calculations are not in agreement with the simulations of Ref. [6]. It is because of the difference in the way of calculating the Hamiltonian density.

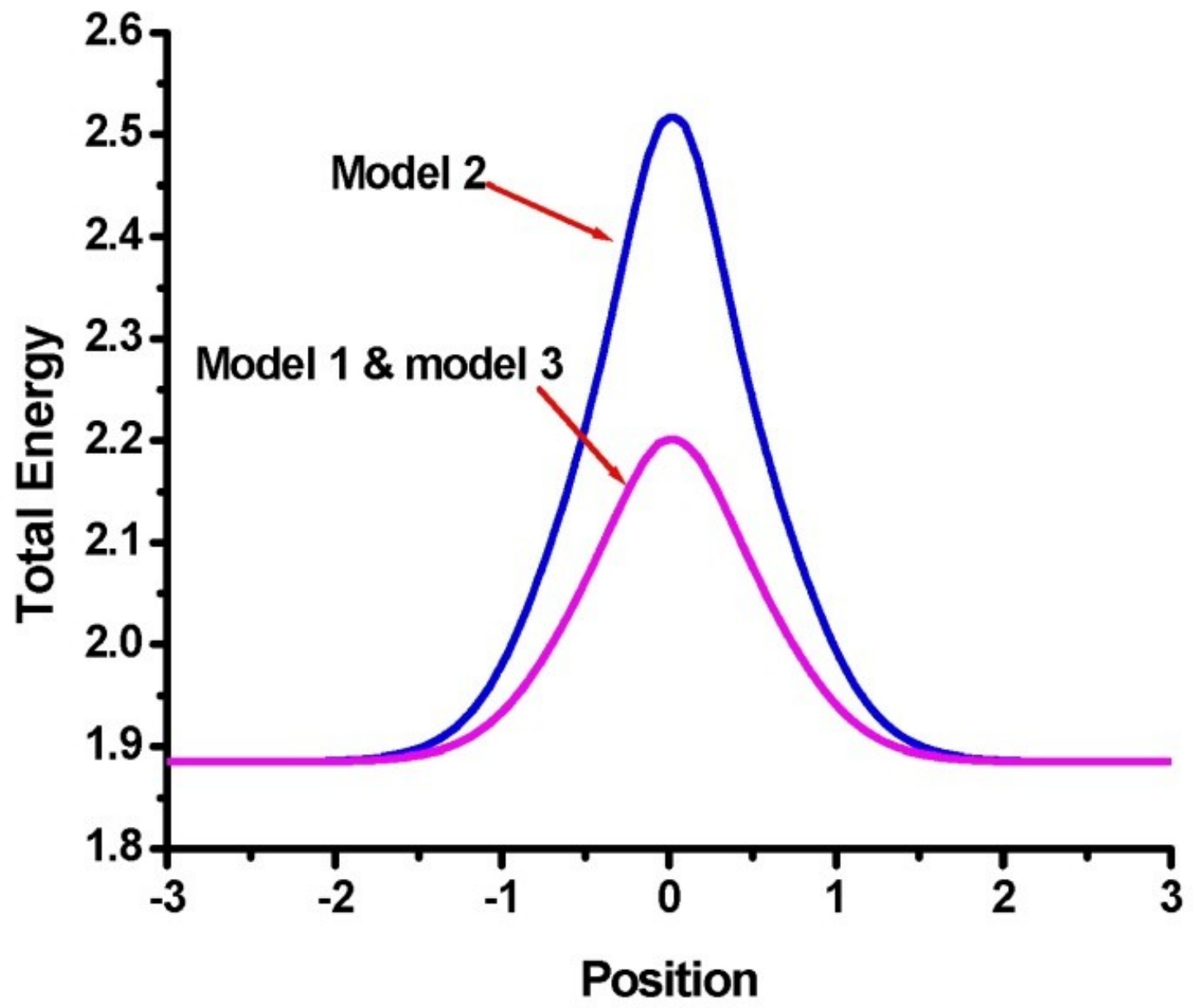

Figure 1: The potential barrier $v(x) = 0.5\ Exp(-4x^2)$as seen by the soliton in three models.

The critical velocity for the soliton to pass over the potential barrier has been demonstrated as a function of the barrier height in figure 2 for three models. There are some differences between the results of the models. The results of models 1 and 3 are more similar and they are different from the results of the model 2.

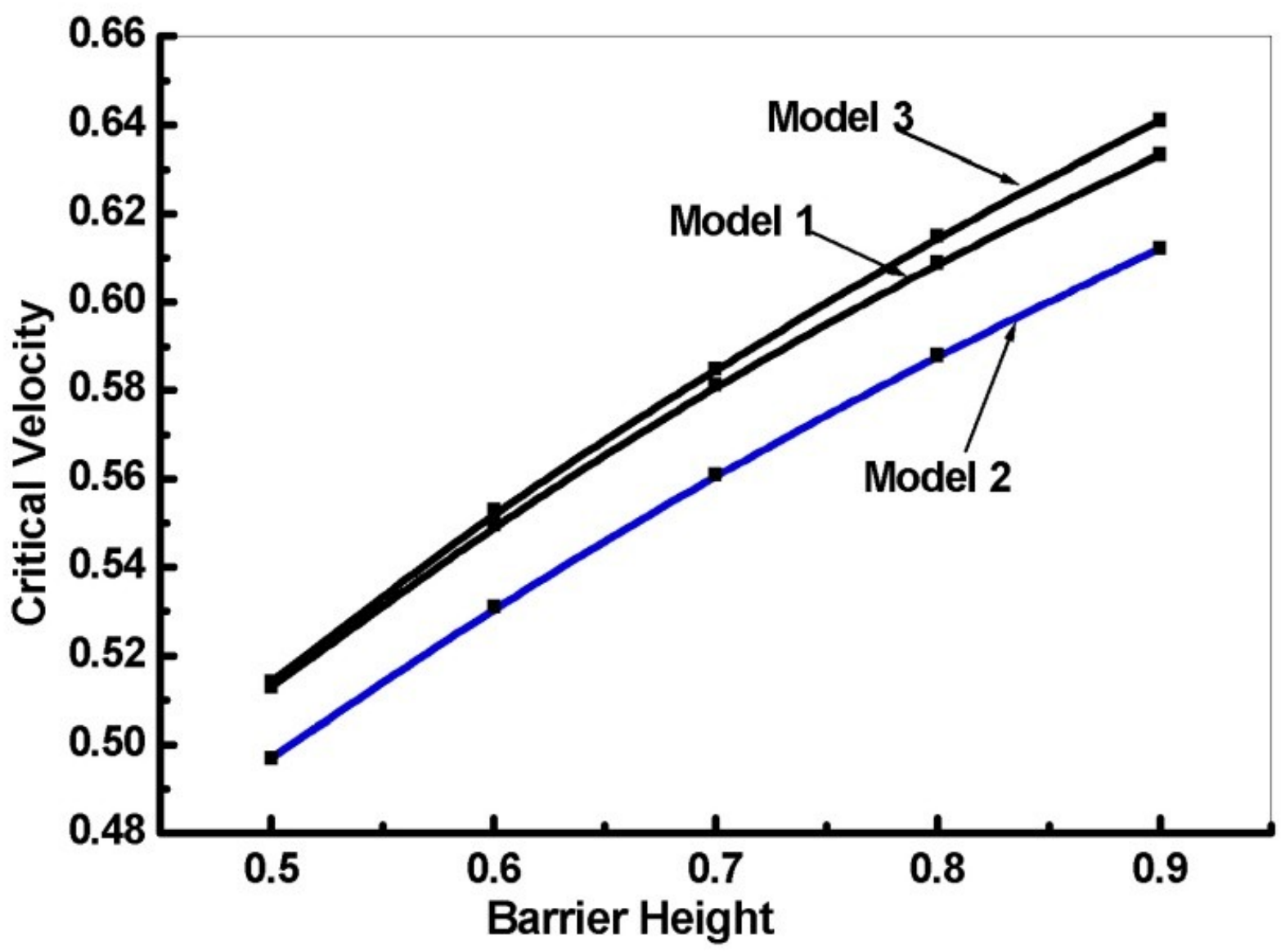

Figure2: Critical velocity as a function of the barrier height of the potential $v(x) = a\ Exp(-4x^2)$ in three models.

Figure 3 shows the critical velocity as a function of the potential width for three models. This figure shows a very good agreement between models 1 and 3 too.

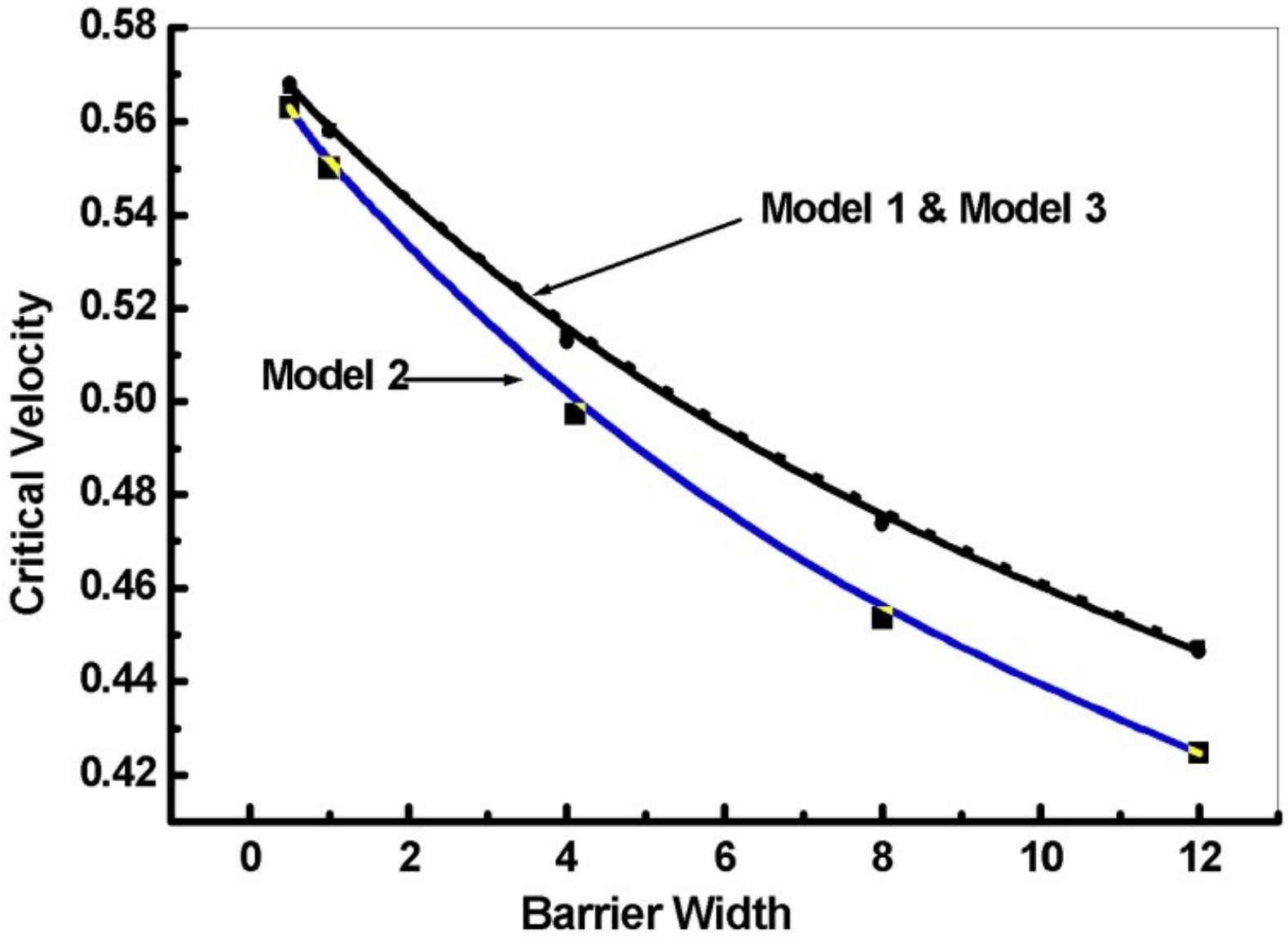

Figure 3: Critical velocity as a function of the barrier width. The potential is $v(x) = 0.5\ Exp(-bx^2)$.

It is mentioned that the static energy of a soliton calculated using the models 1 and 3 are less than the calculated energy in the model 2. Simulations also show that the static energy of the soliton at the peak of the barrier calculated with the models 1 and 3 is the same and they are less than the static energy calculated using model 2 (See figure 1). Figures 2 and 3 show that the critical velocity simulated using the models 1 and 3 are greater than the critical energy in the model 2.

If we look at the soliton as a point particle, we can find the critical velocity with comparing the soliton energy in different positions. In this manner, we expect to find a greater critical velocity using the model 2, because the static energy (rest mass) on top of the barrier calculated with the model 2 is greater than the static energy (rest mass) in the other models. But the critical velocity in the model 2 is smaller than the critical velocity of the other models (see figures 2 and 3). It is in contradiction with the results of the figure 1. Note that the figure 1 presents the energy of a static soliton while the critical velocity is a dynamical parameter. An explanation in base of the effective mass has been presented in Ref. [4]. The reasoning can be completed if we inspect the problem by the collective coordinate approach which has been used for the sine-Gordon model previously [7]. The critical velocity is minimum required velocity for a soliton at the initial position of infinity in which the soliton be able to pass over the barrier after the interaction. The soliton energy in the position $X(t) = x_0 - ut$ is $E\big(X(t)\big) = \int_{-\infty}^{+\infty} \mathcal{H}\big(x, X(t)\big)dx = \frac{M}{\sqrt{1-u^2}}$ where M is the soliton rest mass. The Hamiltonian density of each model is calculated by inserting the solution 4 in the equations (9), (13) and (14) for the models 2, 3 and 1 respectively. Figure 4 presents the rest mass of the soliton as a function of the Barrier height. This figure shows that the rest mass of the soliton in the model 2 is greater than the soliton rest mass in the models 1 and 3. Also models 1 and 3 predict almost the same rest mass for the soliton. It is clear that a soliton with a greater rest mass needs smaller velocity to reach a potential peak. This is true for small potentials as we can find in the figure 2. But for greater potentials the differences between the models 1 and 3 become bigger. It is not surprising, because the models are different. But it is possible to fit the results of the models with the definition of an effective potential [7].

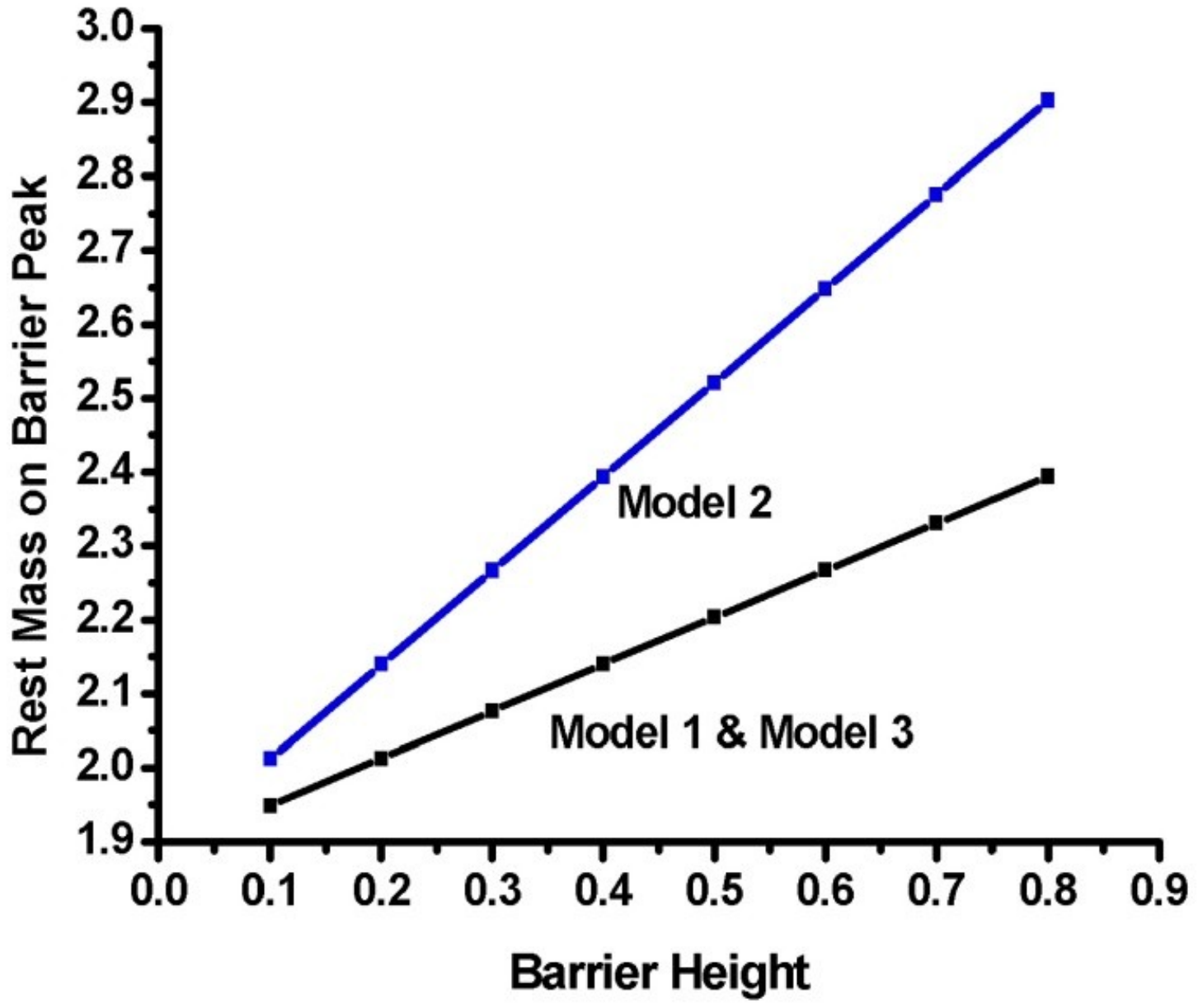

Figure 4: Rest mass of a static soliton on top of the barrier $v(x) = a\ Exp(-4x^2)$ as a function of the height of the barrier

Scattering of topological Solitons on a potential well is more interesting. Unlike a classical point particle which always transmits trough a potential well, a soliton may be trapped in a well if the potential well has enough depth. We have found from the figures 1 and 4 that a soliton has not a fixed mass, so we can not look at the soliton as a point particle in some cases. Several simulations have been done for soliton-well system using three models. Like the potential barrier, the general behaviour of the system is almost the same for all three models. But there are some differences in the details of the interactions. Figure 5 presents a comparison between the shapes of the potential well $v(x) = -0.5\ Exp(-4x^2)$ as seen by the soliton in three models. Models 1 and 3 provide very similar potential but the shape of the potential in model 2 is different.

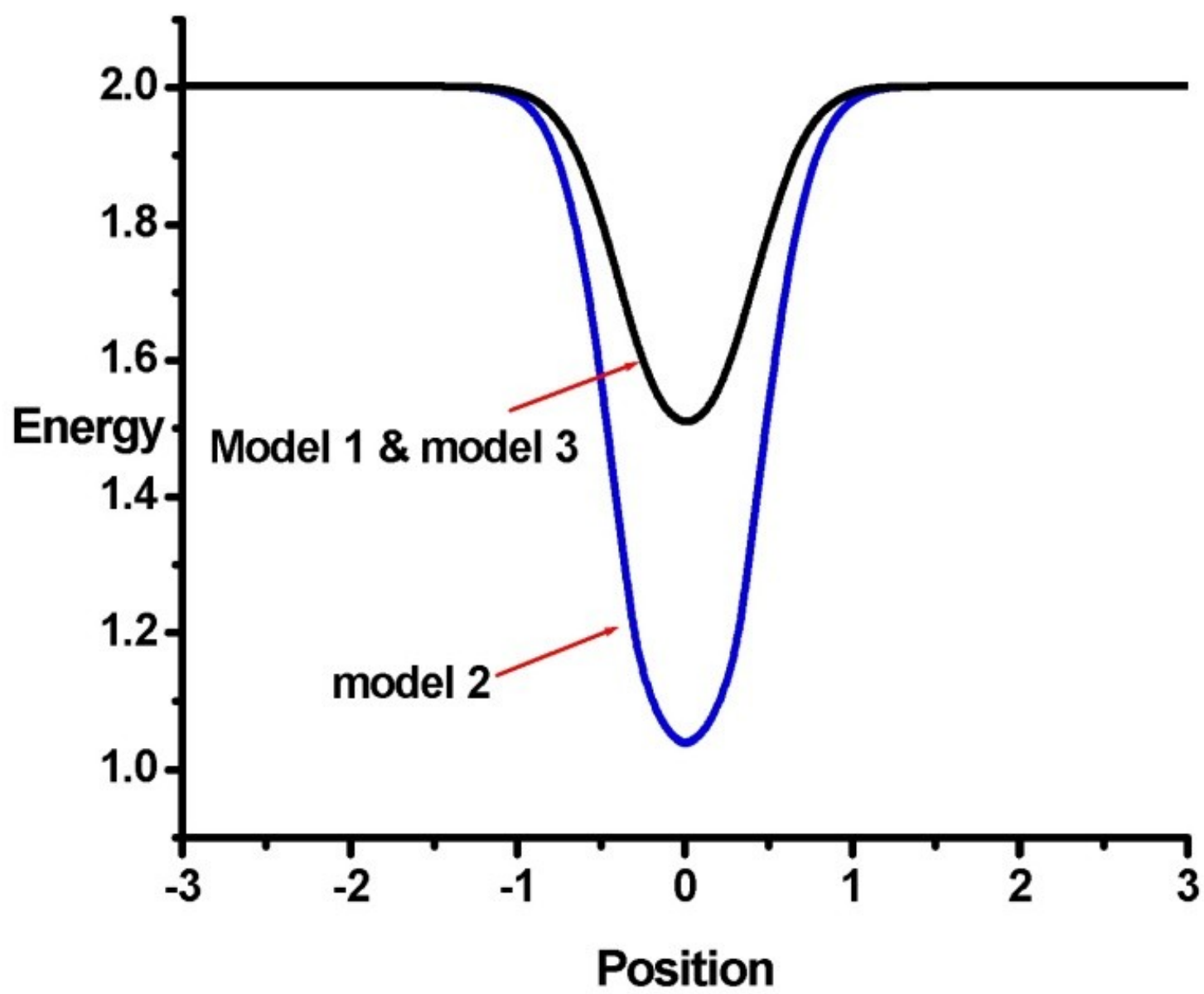

Figure 5: The shape of the potential well as seen by the soliton for the potential $v(x) = -0.5\ Exp(-4x^2)$ .

The differences in the effective potential for the three models cause some differences in the characteristics of the system. For example, the rest mass of the soliton is different when it is calculated using different models. Figure 6 demonstrates the rest mass of the soliton in three models as a function of the potential depth. Models 1 and 3 predict very similar values which are a little greater than the effective mass of the soliton in the model 2.

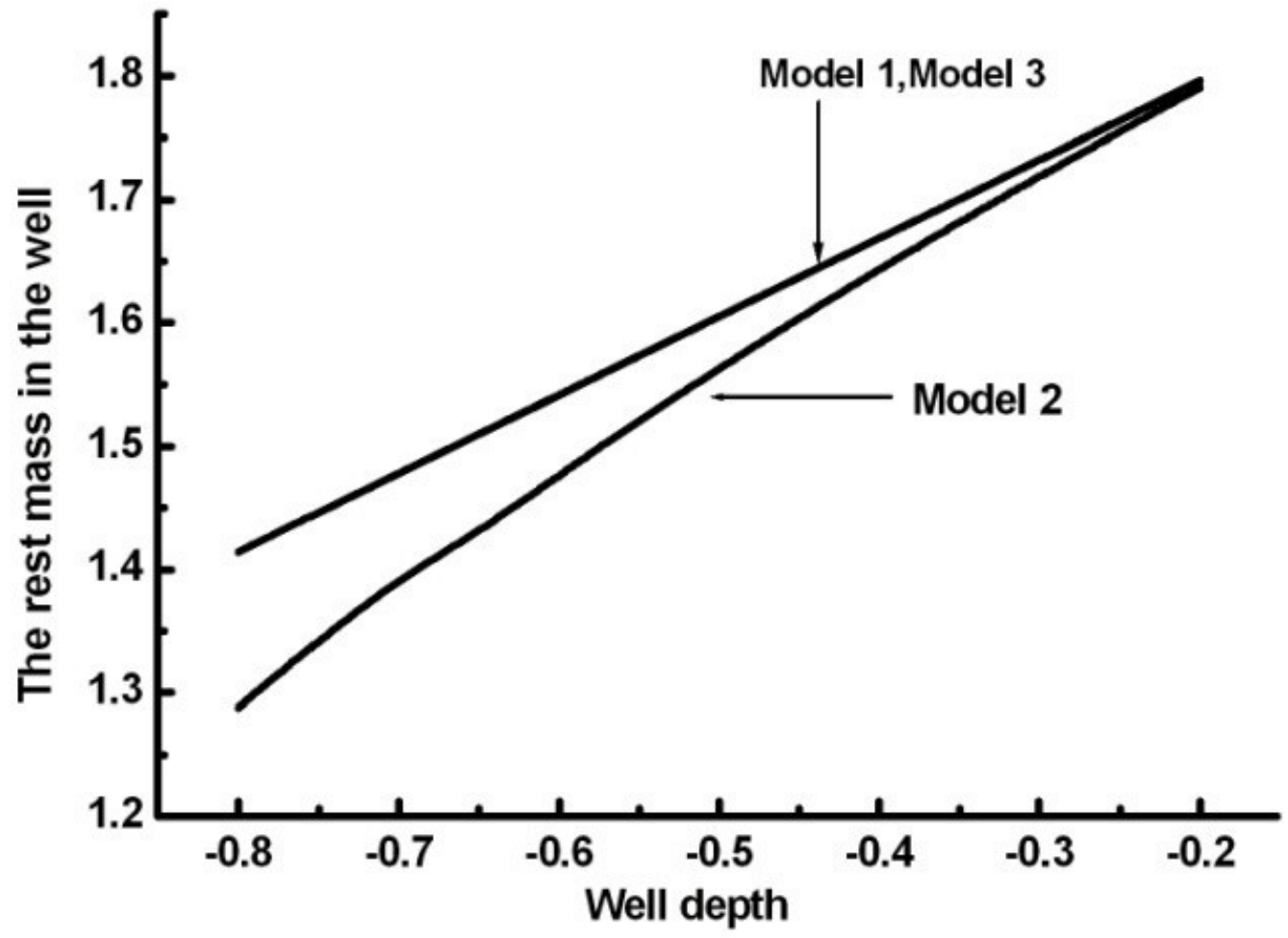

Figure 6: The rest mass of the soliton in the potential well calculated with three models.

As the figure 5 shows, the effective potential in the models 1 and 3 are greater than the effective potential calculated with model 2. On the other hand, the figure 6 indicates that the rest mass of the soliton in the model 2 is less than the rest mass in the other models. Therefore it is expected that the critical velocity of a soliton in the model 2 becomes greater than the critical velocity in the models 1 and 3. Figure 7 presents the critical velocity of a soliton in the potential well as a function of the potential depth. However the potential in the models 1 and 3 is almost the same, but the critical velocity in these models are not equal. It is interesting to compare the soliton trajectory in three models.

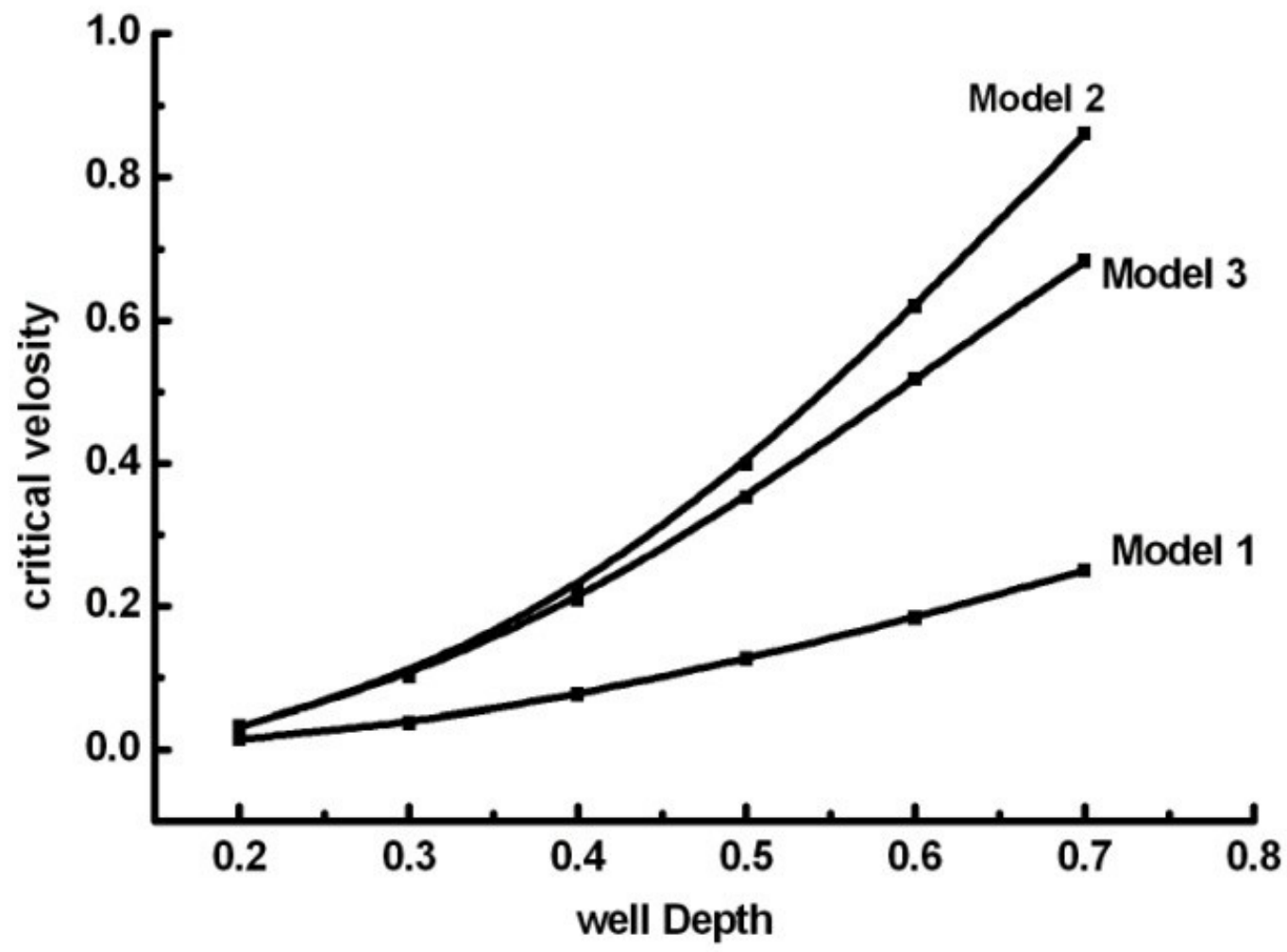

Figure 7: Critical velocity as a function of the depth of the potential well.

Figure 8 presents the trajectory of a soliton with an initial velocity u=0.4 during the interaction with potential $v(x) = -0.5\,Exp(-4x^2)$. The final velocity of the soliton in the model 2 is smaller than the other two models. This means that the energy radiation in the model 2 is very greater than the models 1 and 3. Figure 8 also show that the energy radiation in the model 3 is greater than the model 1. It is the main reason for the differences between the critical velocities in the figure 7.

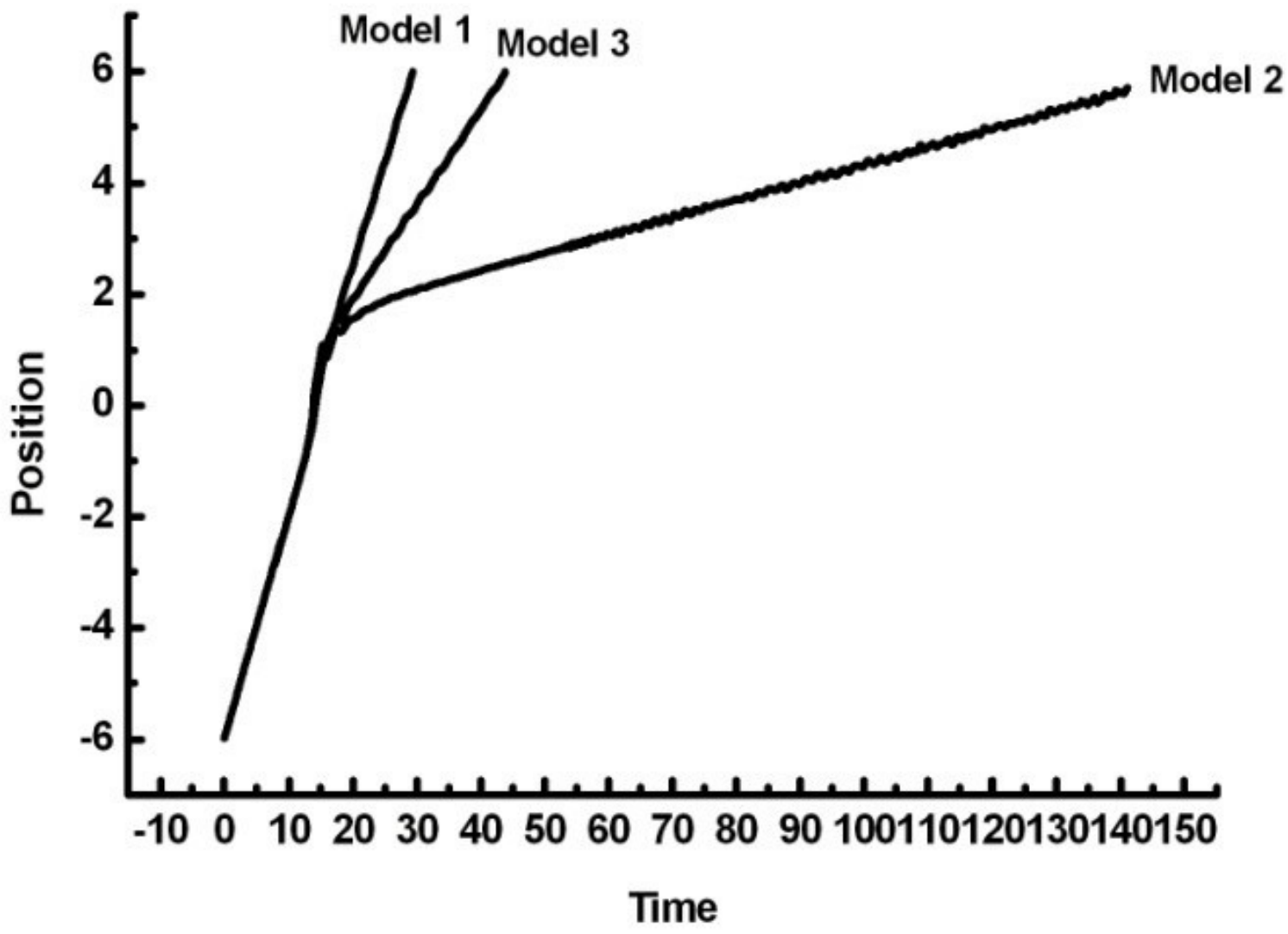

Figure 8: Trajectory of a soliton with initial velocity u=0.4 during the interaction with the potential well $v(x) = -0.5\ Exp(-4x^2)$ in three models.

The most interesting behaviour of a soliton during the scattering on a potential well is seen in some very narrow windows of initial velocities. At some velocities smaller than the $u_c$ the soliton may reflect back or transmit over the potential while one would expect that the soliton should trap in the potential well. These narrow windows can be found by scanning the soliton initial velocity with small steps. Figures 9 show this phenomenon in the model 1 and model 3. Figure 10a shows that a soliton with an initial velocity in the window $0.03 \leq u_i \leq 0.03175$ reflects back during the interaction with potential $v(x) = -0.2\ Exp(-4x^2)$ simulated using the model 3. Figure 10b presents the same phenomenon as in model 2 for the potential $v(x) = -0.6\ Exp(-4x^2)$. Same situations have not been reported in [4] for the model 1. Models 2 and 3 are built by varying the lagrangian density with respect to both "field" and the "metric". Therefore energy exchange between the field and the potential in these models is possible. Soliton reflection in a potential well is a result of energy exchange between the soliton and the potential [1]. This phenomenon needs deeper investigations.

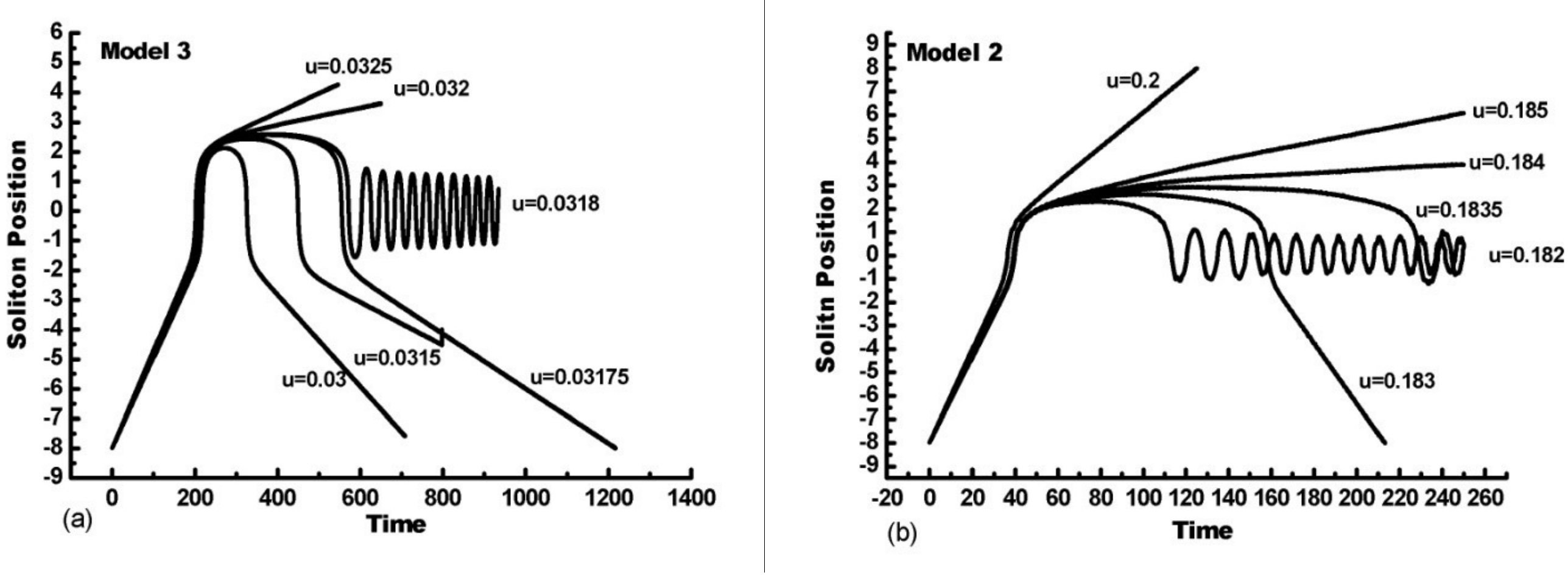

Figure 9: (a) Soliton reflection from the potential well$v(x) = -0.2\ Exp(-4x^2)$ in model 3. (b) Soliton reflection from the potential well $v(x) = -0.6\ Exp(-4x^2)$ using the model 2 .

## 5. Conclusion and remarks

The presented new model is compared with other two models. The results of the interaction of a soliton with potentials using three models are almost in agreement with each others. Model 1 adds the potential to the equation of motion by a different method from those are used in models 2 and 3. Model 3 has better approximation than the model 2. On the other hand model 3 predicts the characteristics of the system very near to predictions of the model 1.

Therefore it can be concluded that the results of the models are valid. The model 1 is easy to simulation while the model 3 is more analytic.

All three models agree that the interaction of a soliton with a potential barrier is nearly elastic. At low velocities it reflects back but with a high velocity climbs the barrier and transmits over the potential. Soliton radiates some amounts of energy during the interaction with the potential. There exists a critical velocity which separates these two kinds of trajectories. Interaction of a soliton with potential well is more inelastic. It is possible that a soliton scatter on a potential well and reflect back from the potential. The models 2 and 3 predict this behaviour.

The model 3 works better than the model 2. Therefore the constructed collecting coordinates using the model 3 may give us better information about the soliton dynamics.

It is interesting to investigate scattering of solitons of other models on defects using model 3. These studies help us to find better knowledge about the general behaviour of solitons.